\documentclass[letterpaper, 10 pt, conference]{ieeeconf}  % Comment this line out
\IEEEoverridecommandlockouts                              % This command is only
\usepackage{graphics} % for pdf, bitmapped graphics files
\usepackage{epsfig} % for postscript graphics files
\usepackage{mathptmx} % assumes new font selection scheme installed
\usepackage{times} % assumes new font selection scheme installed
\usepackage{amsmath} % assumes amsmath package installed
\usepackage{amssymb}  % assumes amsmath package installed
\usepackage{subfigure}

\title{\LARGE \bf
CMOS-Memristor Hybrid Integrated Pixel Sensors
}

\author{ \parbox{6 in}{\centering  Kamilya Smagulova, Aigerim Tankimanova and Alex Pappachen James
        \\
        Bioinspired Microelectronics Systems Group,\\School of Engineering,
        Nazarbayev University, Astana\\
        {\tt\small  www.biomicrosystems.info}}        
}

\begin{document}

\maketitle
%\thispagestyle{empty}
% * <kamilya.smagulova@gmail.com> 2016-07-16T17:29:28.837Z:
%
% ^.
%\pagestyle{empty}

%%%%%%%%%%%%%%%%%%%%%%%%%%%%%%%%%%%%%%%%%%%%%%%%%%%%%%%%%%%%%%%%%%%%%%%%%%%%%%%%
\begin{abstract}
Increase in image resolution require the ability of image sensors to pack an increased number of circuit components in a given area. On the the other hand a high speed processing of signals from the sensors require the ability of pixel to carry out pixel parallel operations. In the paper, we propose a modified 3T and 4T CMOS wide dynamic range pixels, which we refer as 2T-M and 3T-M configurations, comprising of MOSFETS and memristors. The low leakage currents and low area of memristors helps to achieve the objective of reducing the area, while the possibility to create arrays of memristors and MOSFETs across different layers within the chip, ensure the possibility to scale the circuit architecture.  
%This electronic document is a ÒliveÓ template. The various components of your paper [title, text, heads, etc.] are already defined on the style sheet, as illustrated by the portions given in this document.

\end{abstract}

%%%%%%%%%%%%%%%%%%%%%%%%%%%%%%%%%%%%%%%%%%%%%%%%%%%%%%%%%%%%%%%%%%%%%%%%%%%%%%%%
\section{Introduction}

Growing number of applications that involve image sensors leads to a demand for cameras with wide dynamic range and that are compatible with existing technologies. Dynamic range (DR) is an indicator that shows the ability of digital camera to reproduce the captured scene. Rapid development of Complimentary-Metal-Oxide-Semiconductor (CMOS) technology and cheaper manufacturing costs make CMOS image sensors dominant over Charge-Coupled devices (CCD) technology \cite{c1}. The simplest traditional CMOS three-transistor (3T) pixels architectures have linear response that limits their dynamic range to 3-4 decades of illumination whereas human eye is capable of capturing scenes of 100dB \cite{c2}. Thus, numerous other configurations to improve dynamic range of sensors have been proposed.  Particularly using logarithmic response pixels allow capturing scenes with DR more than 5-6 decades \cite{c3}. With the increase in the requirements to have high resolution images, and increased emphasis on high speed processing, the demands to pack more pixels in a unit area of chip has increased over a period of years.

The memristor is characterized by relationship between the electric charge $q$ and the magnetic-flux \cite{c4}. A popular implementation of the programmable two terminal memristor was introduced by the HP Lab which has two resistive states. The device structure represents an oxide TiOx where one part is an insulator TiO2 and the other is oxygen-poor TiO2-x enclosed between two platinum terminals \cite{c5}.  Currently memristor is being recognized as a promising element in analog circuits for various purposes \cite{c6, c6a}. The low area of chip, ease of programmability, the ability to act as a resistive memory in a crossbar array, and low leakage currents make the memristors an attractive device.

In this paper, explore a possible new application of memristor in the design of pixel sensor. Two different configurations of the CMOS-memristor integrated pixels by the analogy of a conventional 3T/4T logarithmic pixel and linear-logarithmic response pixel is proposed. 

%\section{Background}

\section{Proposed Circuits}

Traditional pixels have linear response that limits the dynamic range. One of the ways to extend its DR is utilization of logarithmic output pixels. There are variety of architectures that allow to obtain non-linear responses.

\subsection{2T-M Pixel Design}

\begin{figure}[ht]
\centering
\includegraphics[width=65mm]{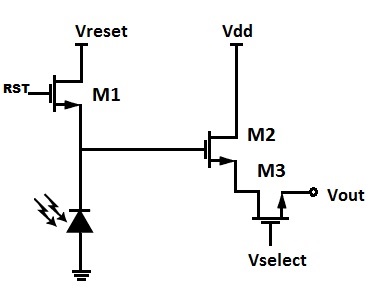}
\caption{ 3T logarithmic pixel configuration.}
\label{f1}
\end{figure}

Fig. ~\ref{f1} represents an architecture of a typical 3T logarithmic response pixel sensor.  
%In this pixel, NMOS transistor M1 between a photodiode PD and a supply voltage $V_{dd}$ serves as a load transistor which works in subthreshold region. ***

%The transistors M2 and M3 act as source-follower select circuit that allow the pixel output to be read by the read-out circuit.

\begin{figure}[ht]
\centering
\includegraphics[width=65mm]{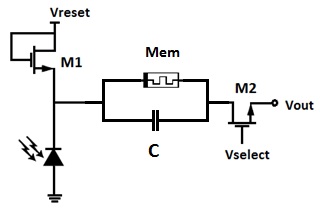}
\caption{Architecture of the proposed CMOS-memristor logarithmic pixel}
\label{f2}
\end{figure}

Leakage currents in transistors and noises that occur due to mismatch of CMOS elements in the sensor circuits severely degrade  the quality of obtained images.The Source follower is formed of NMOS transistor and is utilized to amplify the signal which also amplifies the existing noises and additional ones. To decrease its influence on the quality of output signal we propose to replace transistor M2 with parallel circuit formed of memristor $Mem$ and capacitance $C$. The suggested configurations of the novel pixel is provided in the Fig.~\ref{f2}.

\begin{figure}[ht]
\centering
\includegraphics[width=80mm]{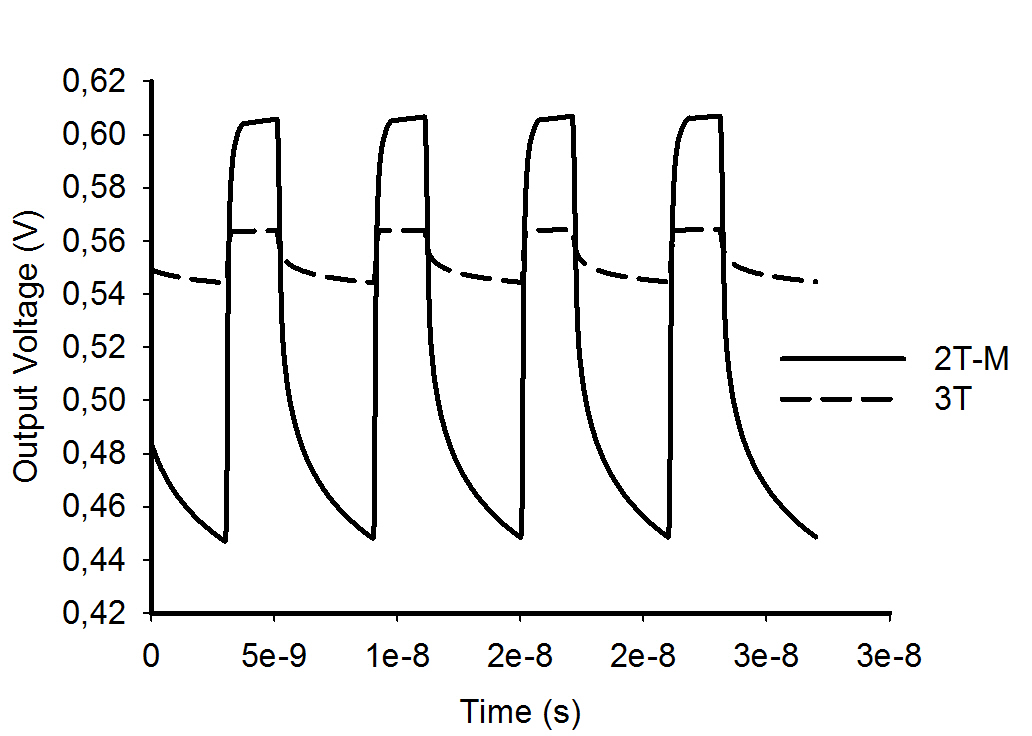}
\caption{Transient analysis of a 3T logarithmic pixel and CMOS/memristor pixel.}
\label{f3}
\end{figure}

The circuit was designed using 90 \textbf{n}m IBM technology and 40nm$\times$90nm TiO$_2$ memristor \cite{c11}. The simulated transient outputs of the 3T logarithmic pixel sensor ($V_{out2}$) and the proposed CMOS-memristor logarithmic pixel sensor ($V_{out1}$) are provided in the Fig. ~\ref{f3}. As it can be seen from the figure, the circuit with memristor demonstrates better performance in terms of signal amplification. The output amplitude range of the 3T logarithmic circuit is 0.2 Volts, whereas in the novel circuit it is twice larger and equal to 0.4 Volts.

Replacement of a source follower NMOS transistor M2 in the Fig.~\ref{f1} with a memristor device as it is shown in the  Fig~\ref{f2} can benefit in various ways without degrading the image quality. Memristor-based cross-bar architectures can be fabricated directly above the CMOS circuit. Utilization of memristors improve scalability of the circuit ~\cite{c7,c8}.
 In addition it can allow execution  parallel operations by the pixel circuit.  Typically memristor size which can significantly decrease the pixel area and as a result can decreases a fill factor coefficient. Further, memristors has lower power consumption compare to MOSFET transistors due to their non-volatile nature \cite{c9}.

% \begin{table}
% \centering
% \caption{\label{tab:widgets} Comparison of configurations.}
% \begin{tabular}{l||c||r} \hline
% Pixel Type & Area, $mm_{}$ & Power Consumption, W \\\hline
%  3T logarithmic pixel  & 42 & ? \\
% CMOS-memristor pixel & 13 & ?\\\hline
% \end{tabular}
% \end{table}

\subsection{3T-M Pixel Design}

\begin{figure}[ht]
\centering
\includegraphics[width=80mm]{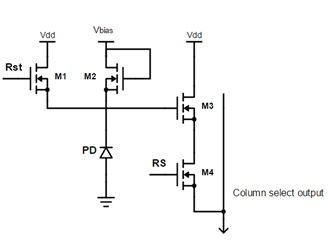}
\caption{ Logarithmic linear response CMOS pixel}
\label{f4}
\end{figure}

\begin{figure}[ht]
\centering
\includegraphics[width=80mm]{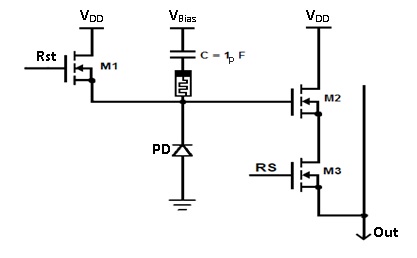}
\caption{CMOS-memristor based linear-logarithmic pixel.}
\label{f5}
\end{figure}

The combined linear and logarithmic response was also a topic of active investigation as it promised increased dynamic and better performance. The combined linear-logarithmic pixel uses a weak inversion mode of transistor \cite{c10}. The output response of such sensors split in two levels of operation such that until some photocurrent level the output voltage change is linear and after some level of input current it changes to logarithmic behavior. 
Our CMOS-Memristor based pixel was designed by the analogy to the 4-transistor linear-logarithmic pixel in the Fig~\ref{f4}  \cite{c10} that was previously discussed. The main idea was to replace the transistor M2 (Fig~\ref{f4}) which operates in weak inversion mode by equivalently operating memristor and capacitor combined in series. By applying this technique, we can replace transistor by elements of smaller size and reduce the total chip area. Capacitance in series with memristor can be of a very small value. Capacitor with memristor in series operates very similar to the transistor in weak conversion mode. In our case, we have chosen the capacitance of 1 pF value. The proposed circuit is shown in the Fig~\ref{f5}.

\begin{figure}[ht]
\centering
\subfigure[]{
\includegraphics[height=30mm]{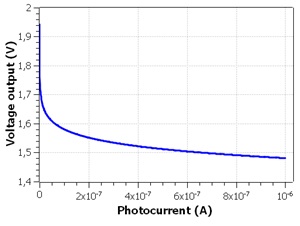}
\label{}}\subfigure[]{
\includegraphics[height=30mm]{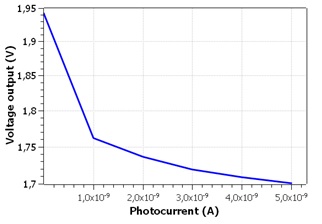}
}
\caption{a) Output voltage of pixel circuit versus input photocurrent; b) Linear-logarithmic response of the proposed CMOS/memristor design.}
\label{f6}
\end{figure}

\begin{figure}[ht]
\centering
\includegraphics[width=80mm]{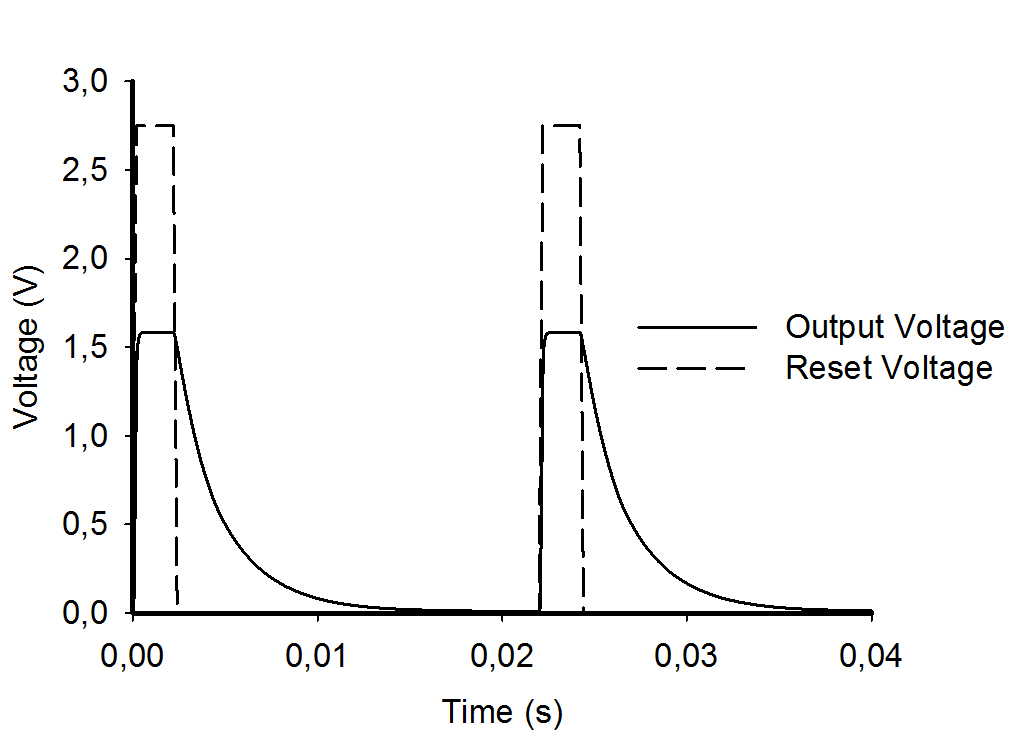}
\caption{4T CMOS/memristor pixel operation cycle.}
\label{f7}
\end{figure}

The pixel output voltage to different photocurrent levels is presented on Fig~\ref{f6} (a). The logarithmic response can be observed in the range between 10$^{-9}$ and 10$^{-6}$ A. The linear response occurs for low photocurrent values, as it can be observed on Fig~\ref{f6} (b).
Fig~\ref{f7} shows results of transient analysis of the pixel circuit is shown. During the reset value is high the output is driven to high voltage and photo diode is charging. When the reset is in off state, the read out period starts and the accumulated charge is transferred from the pixel area to the interfacing circuits.

\begin{figure}[ht]
\centering
\includegraphics[width=80mm]{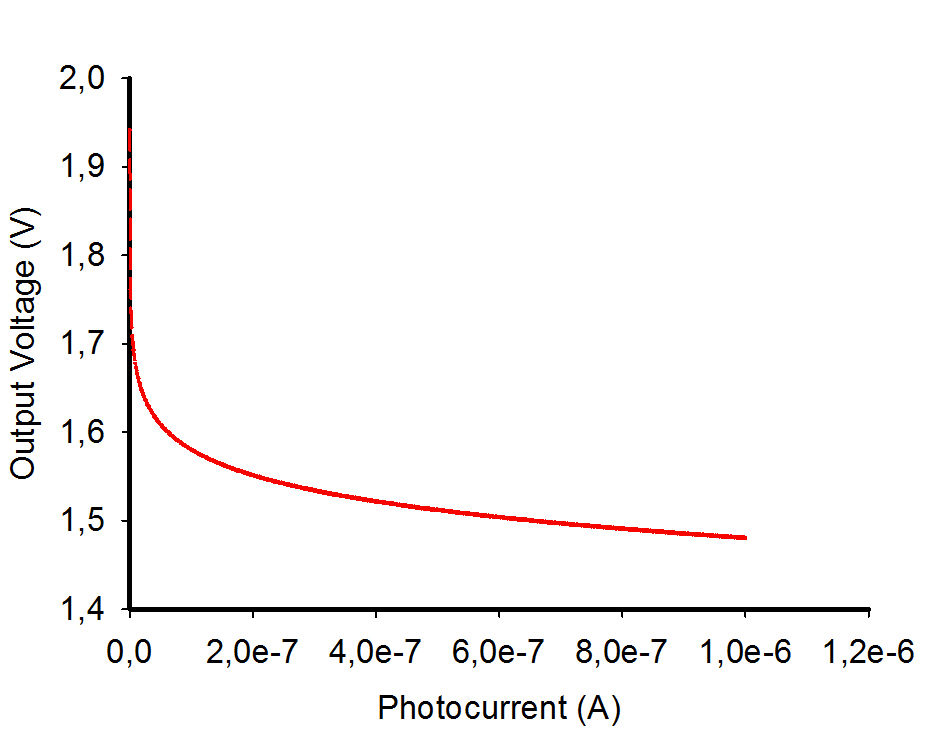}
\caption{Simulation results of 4T CMOS linear-logarithmic pixel response}
\label{f8}
\end{figure}

\begin{figure}[ht]
\centering
\includegraphics[width=80mm]{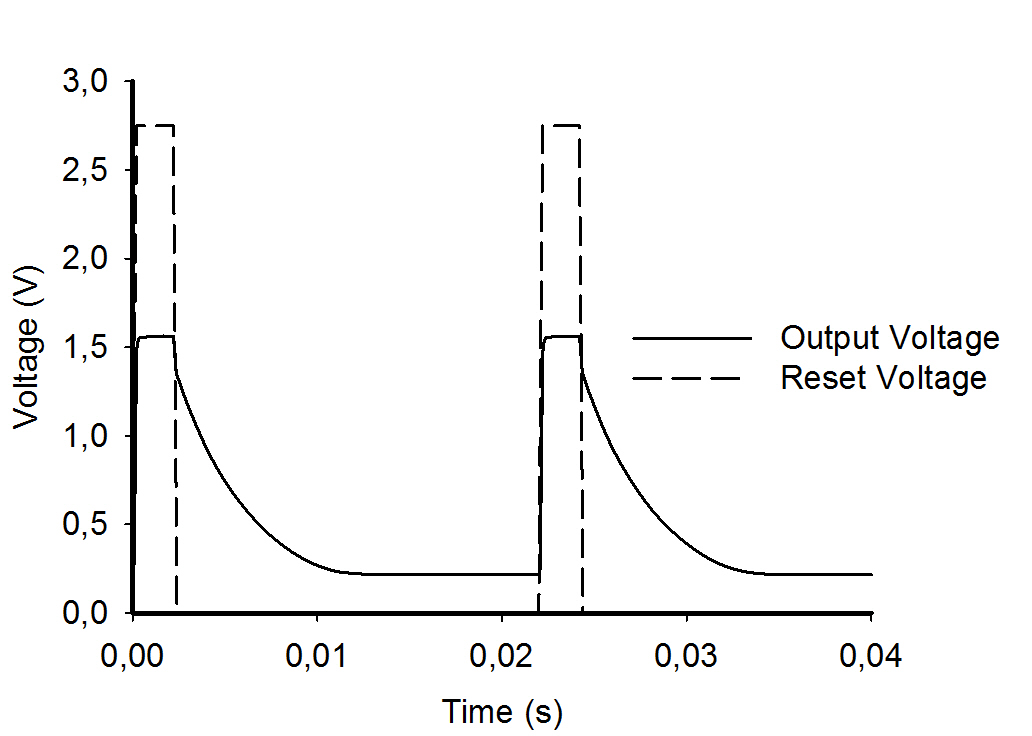}
\caption{Operation cycle of 4T CMOS linear-logarithmic pixel response.}
\label{f9}
\end{figure}

\begin{table}
\centering
\caption{\label{tab:widgets} Comparison of 3T-M with 4T pixel}
\begin{tabular}{l||c||c} \hline
Pixel Type & Area, pm$^{2}$ & Power Consumption, mW \\\hline
 3T-M pixel  & 26.83 & 0.00376 \\
4T CMOS pixel & 100.00 & 0.000605\\\hline
\end{tabular}
\end{table}

Table 1 shows the comparison of the proposed 3T-M pixel with that of 4T CMOS pixel in terms of the area and power. The values noted are the worst case analysis in terms of the power, switching the memristor state to higher resistive state can reduce the power in the 3T-M pixel. On the other hand, there is a substantial reduction in the on chip area with such implementation. This can be further reduced by using a crossbar architecture for placing the memristors over a layer of 3T cell arrays.

Fig.~\ref{f8} and Fig.~\ref{f9} present simulation results for 4T CMOS pixel sensor that was shown on Fig.~\ref{f4}. As it is seen from the graphs on Fig.~\ref{f6} and Fig.~\ref{f9}, the response of the proposed design is similar to the original pixel with the transistor in weak inversion. We achieved the same linear-logarithmic behavior by inserting memristor and capacitor. It is also noticeable that the dynamic range of the proposed pixel is comparably the same as of the 4T CMOS pixel which makes it a good alternative for high dynamic range imager.  
\section{Conclusion}

In this paper we explore the possibility to integrate memristor devices into the CMOS pixel array with an aim to reduce the area without degrading the dynamic range of the pixels. There are several architectures possible with the array configurations such as the memristors in a crossbar architecture can be utlised to build 3D pixel structures, by separating the CMOS and memristor as two separate yet programmable circuits. The ability to program the memristors can also be used to record the state of the pixel in a given time, and can be used for noise correction and further signal analysis problems. 

% The pixels can be further developed to integrate co-processing units such as access to the pixel arrays over network of things.

% The proposed pixel design requires read-out units and other architectural units, that can be developed further to integrate into an internet of things framework. The ability of program memristors and other units within the pixels can help develop cameras that can be optimized with machine learning algorithms remotely.

\addtolength{\textheight}{-12cm}   % This command serves to balance the column lengths

\end{document}